\documentclass[nofootinbib,showpacs,amsmath,twocolumn,showkeys,10pt,aps,pra,longbibliography,floatfix]{revtex4-2}
\usepackage[T1]{fontenc}
\usepackage{inputenc}
\usepackage[margin=1in]{geometry}
\usepackage{color,times}
\usepackage{array}
\usepackage{amsmath}
\usepackage{stackrel}
\usepackage{graphicx}
\usepackage{appendix}
\usepackage{mathrsfs}
\usepackage{amsfonts}
\usepackage{appendix}
\usepackage{color}
\usepackage{comment}
\usepackage{soul}
\soulregister\cite7
\soulregister\ref7
\usepackage{xcolor}
\usepackage [english]{babel}
\usepackage [autostyle, english = american]{csquotes}
\MakeInnerQuote{"}
\definecolor{med-blue}{RGB}{25,25,112} 
\allowdisplaybreaks

\newcommand{\ket}[1]{\vert{#1}\rangle}
\newcommand{\bra}[1]{\langle{#1}\vert}

\begin{document}
\title{High Fidelity Single-NV Qubit Quantum State Tomography by Photoelectric Readout}
\author{Boo Carmans$^{1,a}$} \author{Michael Petrov$^{1,a}$}  \author{Milos Nesladek$^{1,2}$}\email{milos.nesladek@uhasselt.be}
\affiliation{
$^{1}$Q-Lab, Hasselt University, Institute for Materials Research (IUMAT), Q-Lab, Martelarenlaan 42, B-3500 Hasselt, Belgium}
\affiliation{$^{2}$imec, IUMAT, Wetenschapspark 1, B-3590 Diepenbeek, Belgium}
\affiliation{Email Address: milos.nesladek@uhasselt.be}

\begin{abstract}

  The nitrogen-vacancy (NV) center in diamond is one of the advanced candidates for solid-state quantum computing at room temperature and has proven to be competitive in terms of gate fidelity and the availability of quantum error correction codes. Ambient solid-state qubits may therefore provide quantum computing architectures as an alternative to current solutions. The NV center has an associated electronic spin that is conventionally read out via photoluminescence. As an alternative, the electrical readout  is not  limited by the confocal  resolution of quantum microscopes, which is required in the development of NV-based quantum processors to establish a scalable individual single-spin readout on the order of a few nanometers (necessary for strong dipole-dipole coupling). In this work, the performance of photoelectric readout versus traditional optical readout is investigated to perform Quantum State Tomography measurements on a single NV center. A state fidelity of 0.995 $\pm$ 0.006  is achieved using photoelectrical spin-state readout for state reconstruction, comparable to simultaneous optical measurements. We report photocurrent signal delays, which significantly slow down the photocurrent tomography. We develop a model describing their origin, and propose strategies for their mitigation.
   
\end{abstract}

\maketitle

\section{Introduction}\label{Intro}
For the implementation of quantum computing, the optimal choice of hardware depends on the concrete application field. \cite{healthcare} \cite{finance} Superconducting, trapped-ion or Rydberg atom systems are well studied and show excellent fidelity. \cite{iontrap_review} \cite{supercond_quantumcomp} \cite{supercond_1qubit} \cite{ion_12qubit} \cite{Rydberg1} \cite{Rydberg2} On the other hand, these systems suffer from a large and costly overhead caused by cooling (cryogenics) or bulky vacuum technology. As an alternative, there are semiconductor spin qubits, e.g. quantum dots or donors in silicon. They constitute a solid-state computing platform with a high potential for technological integration that has demonstrated favorable single qubit gate fidelities \cite{SiFidelity} and large coherence times \cite{3ms_coherence_Qdot} in separate experiments. However, similar to the aforementioned platforms, they require operation at cryogenic temperatures and their scalability is one of the main issues. \cite{Si_review} The development of ambient processors based on solid-state defects in wide bandgap materials such as diamond, silicon carbide, boron nitride and others could significantly accelerate the development of portable quantum applications for use at ambient conditions. The nitrogen-vacancy (NV) center spin qubit can be engineered in the diamond lattice and is one of the leading candidates for room-temperature quantum computing architectures. \cite{diamoncomputing}\\

NV centers are formed by substitutional nitrogen atoms coupled to a first-neighbor carbon vacancy. The defect exists both in the neutral (NV$^0$) and negative charge state (NV$^-$), where the latter shows extensive paramagnetic behavior \cite{NVreview} \cite{reviewShallowspins} and favorable quantum characteristics, such as millisecond ambient spin coherence times \cite{coherence} \cite{coherence2} and room-temperature photostability with well-defined optical transitions \cite{optically_coherent_NV}. The NV$^-$ electron spin triplet is usable for quantum applications due to its bright spin conserving transition, hence NV$^-$ will be referred to simply as "NV center". The electron spin of the NV center can be polarized by optical pumping and read through a technique called Optical Detection of Magnetic Resonance (ODMR). \cite{NVreview} The hyperfine coupling to nearby nuclear spins extends the readout and control to the otherwise hidden nuclear spin dynamics, e.g. the nitrogen nucleus (or other proximal nitrogen atoms) which can be $^{14}$N or $^{15}$N \cite{NV_N_1} \cite{NV_N_2} or $^{13}$C nuclei \cite{NV_13C_1} \cite{NV_13C_2}. This allows to prolong coherence times, as nuclear spins experience a weaker coupling to their environment. By transferring the electron spin state onto the nuclear one, quantum registers  can be created. \cite{QuantumMemory} \\

Important functionalities of a quantum computer have been demonstrated for the NV platform, such as quantum error correction \cite{NV_errorcorrection}, couplings, quantum teleportation \cite{NV_QuantumTeleportation}, etc. For a single NV center, single qubit gates with a fidelity of 0.99995 and two-qubit gates with a fidelity of 0.992 have been demonstrated at low temperatures. \cite{Fidelities_NV} Nearby NV spin qubits can be entangled via their magnetic dipole-dipole interaction (strong coupling regime), which was demonstrated for both the electronic and nuclear spins of an NV pair. \cite{NV_pair_entanglement} NV pairs can be created at such small distances, e.g., by implanting high-energy nitrogen ions into small spots \cite{pair_implant_highE}, by mask techniques \cite{pair_mask} or by molecular implantation of ionized nitrogen molecules. \cite{molecular_implant} A single NV center can also be coupled to nuclear spins, demonstrating a mapping of 27 spins \cite{NVcoupledto27spins} and entangled spins (9) \cite{NV_10spins}. Further up-scaling and improvement of the NV generation rates to create NV spin registers is an important step on the road to the envisioned diamond-based quantum processors. \cite{creation_efficiency}  \cite{75percentyield}\\

Currently, ODMR is used as the main detection method for the NV spin state. This readout method is being employed in spin-photon interfaces, which are used to construct small quantum registers. In these registers, the spin manipulations allow for a spin-selective production of photons, which are then used in photon-entanglement schemes. \cite{spinphoton1} \cite{spinphoton2} In these works, the entanglement rates were rather low, but later entanglement rates have been reached in the range of tens of Hz. \cite{spinphoton3} However, as an alternative, the entanglement generation through NV dipole-dipole coupling has the potential to exceed these rates by orders of magnitude. \cite{dipole1} \cite{dipole2}\\

For reading multiple spin qubits, an efficient scale-up of the technique is required as the optical resolution is limited by the Abbe limit, which is at least an order of magnitude smaller than the interqubit distances that are required for dipolar NV spin entanglement. Secondly, there is the low collection efficiency caused by the high refractive index of diamond and by the design of conventional detection optics. Both problems can be alleviated but only to some extent: the resolution by optical techniques such as STED \cite{STED} (but only for a small number of qubits), the addressability by B-field gradients \cite{Bgradient} and the collection efficiency by microfabrication techniques \cite{microfabrication}. A solution for individual qubit readout is provided by the Photoelectric Detection of Magnetic Resonance (PDMR). \cite{InventionPDMR} This method relies on the direct, on-chip electric detection of charge carriers that are excited from NV centers to the diamond conduction band by a two-photon process under optical pumping. The compatibility of this readout method with coherent manipulations of NV spin ensembles has been demonstrated in ambient conditions \cite{EnsemblePDMR} and the readout contrast has been increased using dual-beam excitation \cite{dualbeamPDMR}. It has also been successfully used to read out a single, coherently manipulated, electronic \cite{singleElectronPDMR} and nuclear \cite{singleNuclearPDMR} spin, demonstrating its potential for qubit state detection. Compared to optical detection, the advantages are spatial resolution (given by the contact size), detection rates, compactness, and integration with electronics. \cite{PDMRreview} The optical-to-electronic conversion of the photoluminescence ($PL$) signal is no longer required and the integration of NV-diamond with electronics is facilitated, paving the way to compact integrated systems.\\

This work extends the photoelectric readout of NV centers to Quantum State Tomography (QST), which includes determining the fidelity of quantum state readout using certain quantum gate operations. In particular, we base ourselves on Rabi Phase Quantum State Tomography (RPQST) \cite{ourQST} and apply it to the electron spin of a single NV center at room temperature. This methodology relies on the phase determination of two orthogonal Rabi measurements of a pure quantum state and has been demonstrated on the electronic and nuclear spin associated to a single NV center in diamond. Here, the technique is adapted to the photoelectric readout and applied to 21 different pure single NV electron spin states (a representative selection for the entire Bloch sphere) and a state fidelity of $0.995 \pm 0.006$ was achieved, which is approximately equal to what was attained by conventional optical detection in \cite{ourQST} (also at room temperature). We demonstrate that the photocurrent ($PC$) readout can be a useful tool to measure spin states with high fidelity and has a potential to be applied in scalable NV spin registers working in the strong dipole-dipole coupling regime. We also discuss the photocurrent readout delay, which we consider to be the main obstacle for realizing the potential advantage of the photocurrent readout. We argue that this readout delay is related to the mechanism of charge collection from diamond.

\section{Materials and Methods} \label{methods}
\subsection{Experimental setup}
\begin{figure*}
  \includegraphics[width=\linewidth]{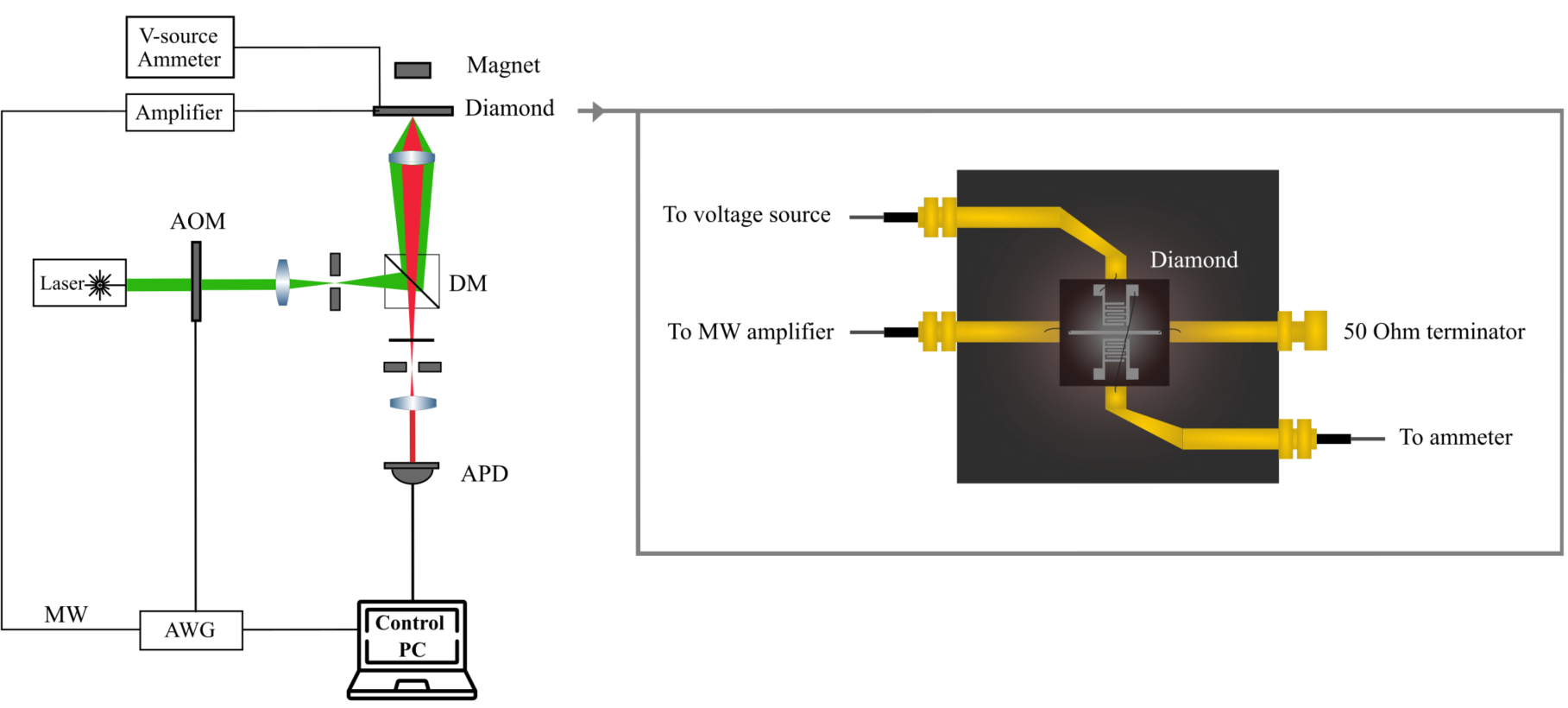}
  \caption{Experimental setup with detail of printed circuit board containing diamond sample and connections to different devices. Electrodes can be seen on top of the diamond sample to collect the charge carriers for the photoelectric readout. }
  \label{experimentalsetup}
\end{figure*}

All measurements were performed on a home-built confocal setup that allows for simultaneous optical and electrical detection and is schematically depicted in \textbf{figure \ref{experimentalsetup}}. The laser excitation originated from a yellow (561 nm) gem laser from Laser Quantum with a continuous-wave power tuned to 3.5 mW. For pulsed operation, the laser was focused onto an acousto-optic modulator (AOM) of type AOMO 3200-146 made by Crystal Technology, Inc. The light passed a dichroic mirror (DM) of type DMLP567R from Thorlabs before being focused onto the sample by an air objective (×40, N.A. 0.95). That same DM allowed fluorescence (637 nm) from the NV centers to pass through towards the detector (a SPCM-AQRH-14 single photon counter from Excelitas Technologies), while blocking the strong laser light. Behind the DM was a FELH05550 Thorlabs long-pass filter with a cutoff wavelength of 550nm and a Thorlabs P30S pinhole to block the out-of-focus light, ensuring that the detected photons mainly originated from the NV center. \\

A high-pressure high-temperature (HPHT) diamond was used as-bought (originating commercially from NDT). This is a type IIA diamond sample of electronic grade with 10 ppb of nitrogen and 50 ppb of boron, a (111) surface and no additional implantation that hosts single defect centers and small clusters. A MW antenna and a set of interdigitated electrodes (3.5 $\mu$m gap) were fabricated onto the diamond surface using optical lithography with sputtering deposition and consisted of a stack of a 20 nm layer of titanium covered with a 100 nm layer of aluminum. For the application of MW pulses, an arbitrary wave generator (AWG) of model M8195A from Keysight was used together with a MW amplifier (ZHL-16W-43-S+, Mini-Circuits). The photocurrent detection used a bias voltage of approximately 1V/$\mu$m delivered by the picoammeter/ voltage source from Keithley (model 487), that was also used for current readout. 

\subsection{Pulse sequence for PC-RPQST methodology}
\begin{figure*}
  \includegraphics[width=\linewidth]{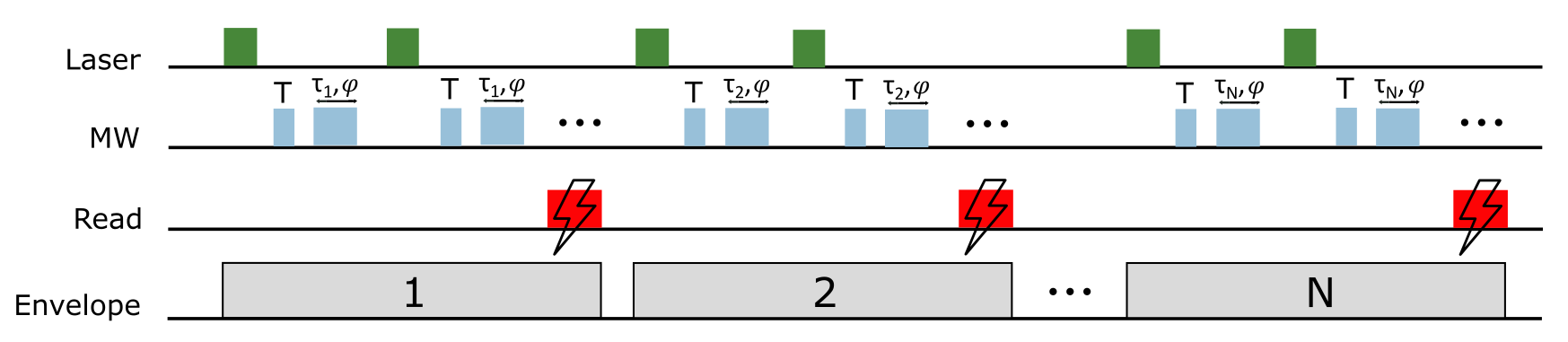}
  \caption{Pulsed photocurrent detection protocol for PC-RPQST. The measurement is divided into envelopes (1, 2, ..., $N$) during which a certain part of the standard pulse protocol is repeated. Each envelope contains the pulse sequence for the initialization of the spin state (always the same) and for one value of the Rabi period $\tau$. Both photocurrent and photoluminescence are recorded at the end of each envelope time $T_E$ to allow for stabilization after each parameter change. After envelope $N$ has been completed, the loop starts again at envelope 1.}
  \label{QST_envelope}
\end{figure*}

This section deals with the quantum state fidelity measurements using RPQST \cite{ourQST} with photocurrent readout (PC-RPQST), a technique that forms the base of our paper and that is depicted in \textbf{figure \ref{QST_envelope}}. There is a caveat associated with the photocurrent detection. Due to the small current values (usually in the range of 1-10 pA for a single NV center), a substantial amplification is needed. However, current amplifiers with sufficient gains (10$^8$V/A) are notoriously slow. This slow measurement speed can be addressed by the envelope methodology, in which a series of fast pulses are integrated into a long envelope (duration = envelope time $T_E$) that is detected by a lock-in amplifier. \cite{singleNuclearPDMR} Here, a picoampere meter is successfully used instead of a lock-in amplifier. This adaptation of standard pulse sequences to envelopes, read by a DC device, can be extended to more complex sequences as PC-RPQST, which relies on the accurate determination of the phase angle for state reconstruction of the electron spin associated to an NV center in diamond. In this case, each of the $N$ envelopes contains the fast RPQST pulse sequence taken from \cite{ourQST} for one set of parameters.

The method starts from the Bloch sphere with the state vector defined as:
\begin{equation}
\ket{\psi} = \cos{\frac{\theta}{2}}\ket{0}+e^{i \phi}\sin{\frac{\theta}{2}}\ket{1}
\end{equation}

Here, $\theta$ and $\phi$ are the polar and the azimuthal angle in the Bloch sphere representation. This state is first constructed, after which Rabi measurements around the x- and y-axis are performed. The related phase angles $\alpha$ and $\beta$ of Rabi oscillations of $\ket{\psi}$ are experimentally determined, after which ($\theta$, $\phi$) can be found by exploitation of the spherical geometry of the Bloch sphere. The full details of the experimental protocol and the relations between the phase angles can be found in \cite{ourQST}. With these phase angles, the density matrix of the NV electron spin $\rho$ can be reconstructed as $\rho_\text{exp}$, such that $\rho_\text{exp}= \ket{\psi_\text{exp}}\bra{\psi_\text{exp}}$. This experimentally determined density matrix is compared to $\rho$. The comparison is performed quantitatively using the state fidelity $F$ \cite{fidelity} as given in formula \ref{eq:F}, which will determine the applicability of the photocurrent readout for this precise, phase-based methodology.

\begin{equation}
F = \frac{\text{Tr}(\rho_\text{th} \rho_\text{exp})}{\sqrt{\text{Tr}(\rho_\text{th} ^2)\text{Tr}(\rho_\text{exp}^2)}}
\label{eq:F}
\end{equation}

\begin{figure*}
  \includegraphics[width=\linewidth]{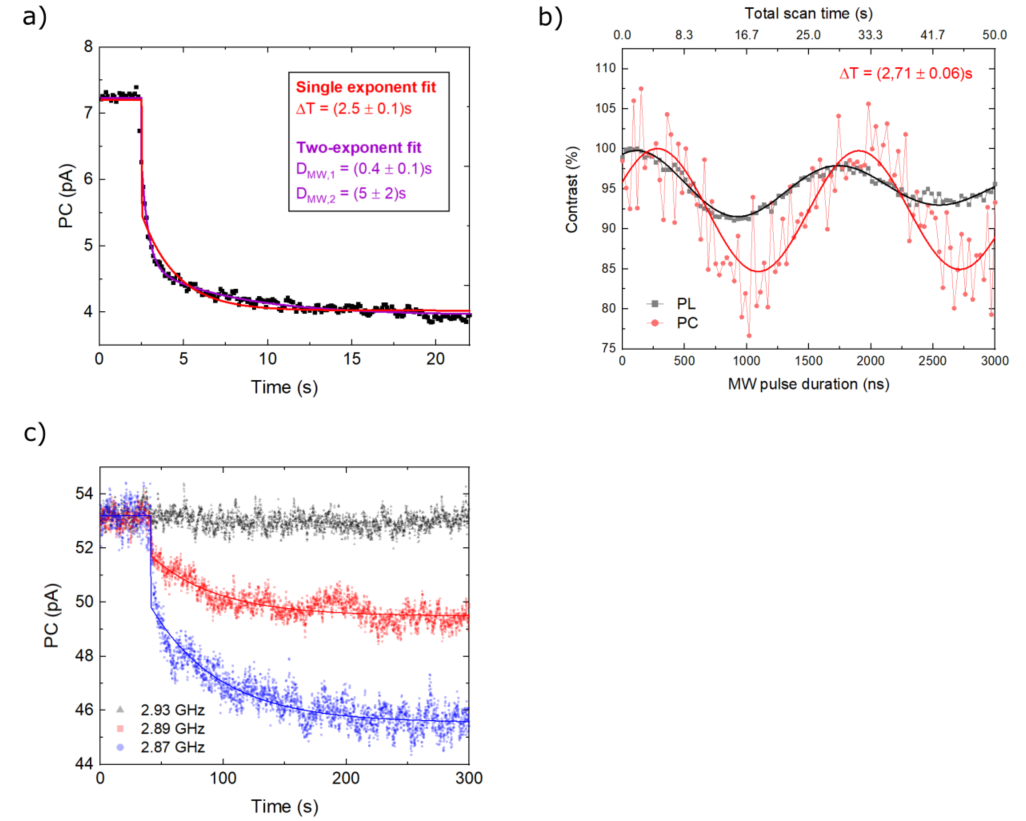}
  \caption{Example of photocurrent delay in \textbf{a)} a $PC$ step response to changing the MW frequency from off-resonance to on-resonance, \textbf{b)} a $PC$-Rabi measurement (red) compared to $PL$ (black) and \textbf{c)} a $PC$ step response to changing the MW frequency from off-resonance (2.77 GHz) to several other frequencies (on-resonance - 2.87 GHZ, slightly detuned - 2.89 GHz and away from resonance - 2.93 GHz) on a different NV center. In \textbf{a)} and \textbf{b)}, the bias voltage is 0V. Note that the $PC$ delay manifests as a phase lag in the Rabi measurement and as an exponential decay in the step response. The time constant $\Delta T$ derived from the Rabi measurement matches approximately the time constant $\Delta T$ from the single-exponent fit (red) of the step response, though the two-exponent fit with time constants $D_{MW,1}$ and $D_{MW,2}$ (purple) follows the data more closely. In \textbf{c)}, the delay constant is approximately equal at different frequency changes. It is different from the value in \textbf{a)} and \textbf{b)} as the NV center was different. }
  \label{link}
\end{figure*}

Importantly, when executing Rabi measurements, a signal phase shift was observed with respect to the optically detected Rabi signal. By further investigation of its origin, a delay in the photocurrent response was detected. \textbf{Figure \ref{link}a)} shows an example of this effect, where the $PC$ responds to a MW edge that switches the frequency from off-resonance to on-resonance. Note, by comparison to \textbf{figure \ref{link}c)}, that the time scale at which the $PC$ changes differs between single NV centers, pointing towards a delay mechanism that is associated with the specific NV center and ruling out hardware issues as origin of the delay times in the range of seconds. The full $PC$ response to a MW edge seems to be fitted best by a two-exponent decay with time constants $D_{MW,1}$ and $D_{MW,2}$. However, a single exponent is used here (time constant $\Delta T$) to describe the system's response at prolonged times, since it can be linked to the phase shift observed in Rabi measurements. The observed phase lag of the PC-Rabi compared to the PL-Rabi is in fact a manifestation of the delay in the $PC$ response which happens under any type of MW driving. Based on the experimental data plotted in \textbf{figure \ref{link}a)} and \textbf{b)}, we consider a mathematical description  of the impulse response function based on Fourier analysis. The derivation of the relation between the MW-step response and the Rabi signal is included in the \textbf{Supplementary Information}. In short, the delay is described by a first order response function with characteristic delay time $\Delta T$, which yields the following relation between the input $i(t)$ and output $o(t)$: $\Delta T \dot o(t) + o(t) = i(t)$. Here, $i(t)$ is related to the instantaneous generation of charge carriers and $o(t)$ to the measured $PC$. A step function input then results in an exponential output and a sinusoidal input in a phase-lagged sinusoidal output. Note that the time constant of the single-exponent decay function in \textbf{figure \ref{link}a)} matches the time constant extracted from the Rabi measurement in \textbf{figure \ref{link}b)}, demonstrating the validity of the first order approximation. In \textbf{figure \ref{link}c)}, an additional measurement is shown, which was performed on a different NV center in the same sample. The response to a switch in MW frequency was measured, going from a starting value of 2.77 GHz to a resonant value of 2.87 GHz, a detuned value of 2.89 GHz and a highly detuned value of 2.93 GHz. The time constants of this delay are approximately equal at 2.87 GHz and 2.89 GHz. At 2.93 GHz, they cannot be determined as the $PC$ contrast is too low. \\

As the correct determination of the Rabi phase angle is essential for PC-RPQST, we paid attention to this phenomenon. Crucially, the delay in the PC-RPQST response (\textbf{figure \ref{link}b)}) is a result of the delay in the photocurrent response (\textbf{figure \ref{link}a)}). Therefore, it can be mitigated by increasing the measurement time per point ($T_E$), which is illustrated in the \textbf{Supplementary Information}. The experimental hardware makes only a small contribution to the delay measured in $PC$, which is mainly caused by the Keithley picoamperemeter. This contribution is limited to the range of 150 ms, which is derived from measurements of the $PC$ response to switching on the laser illumination for multiple NV centers (see \textbf{Supplementary Information}). However, the current delay derived from \textbf{figure \ref{link}} is in the order of 2.5 s and does depend on the specific NV center that is investigated. As the likely reason of these s-scale delay times, there is the complex nature of NV photocurrent. Although the charge generation rate $G$ of the NV center responds immediately to changes in laser or MW driving, this is not always reflected in $PC$ measurements. Here, $G$ refers to the rate at which free electrons and holes are generated by NV center charge cycling, which depends on the NV$^-$ spin state. \cite{PDMRreview}  For example, in \textbf{figure 3c)}, the MW frequency is driven from an off-resonant value to resonance (2.87 GHz - blue curve). Starting from the $m_s$= 0 state, the electron is photo-ionized from the NV$^-$ excited state triplet to the conduction band, while for $m_s$ =$\pm1$ state, the electron is not photo-ionized but undergoes an internal transition to the NV$^-$ ground state. For these two cases, $G$ is different and changes immediately upon switching the MW frequency. However, the $PC$ does not change as fast as $G$, but shows a slower response. From comparing the curves at resonance (blue) and slightly detuned from resonance (red), it can be seen that the amplitude of the initial fast reaction of the $PC$ is larger when the change in $G$ is larger. However, the subsequent slower $PC$ response-time does not depend on MW frequency, but happens at approximately the same time scale for both frequencies.\\

It has been shown that background current, charge carrier recombination or trapping due to other defects can impact both $PC$ size and PDMR contrast \cite{yellow_laser} \cite{Influence_electr} \cite{positivePDMR} and that these effects may be slow \cite{Michael} \cite{chargetrap_electr}. Since the measurements presented here were done on the same diamond sample that was used in \cite{Michael}, we conclude that the charging effects discussed there should be apparent in our measurements as well. These charging effects are happening on the same time scale as the here observed $PC$ delay. Therefore, we speculate that the slow $PC$ response that we observe (and that leads to an apparent phase shift in Rabi measurements) is related to charge trapping in the diamond sample. More details are provided in the \textbf{Supplementary Information}.

\section{Results and discussion}

\subsection{Mitigation of photocurrent delay in PC-RPQST}
As explained in the previous paragraph (and in more detail in the \textbf{Supplementary Information}), the current readout delay translates into a change in the phase of the Rabi measurement detected via $PC$ compared to the optically detected signal. This corrupts the phase information, which leads to a large error on the determination of the final spin state and is thus detrimental for the PC-RPQST state fidelity. \\

To mitigate this effect, one method is to increase the envelope duration $T_E$ until it exceeds the delay time. This removes the phase difference from the Rabi signals, but also increases the measurement time. Alternatively, the delay can be described by a first-order response function and corrected based on a single reference measurement. In this case, we use an envelope duration of 500 ms, which we have found to be the point of optimal trade off between signal-to-noise ratio and measurement time. On a subset of 13 measurements on 7 different electron spin states, we find that this correction method improves the PC-RPQST state fidelity. In particular, it is done by determining the $PC$ delay time from the first measurement and then correcting the subsequent 12 experiments using this value. Doing so, the PC-RPQST state fidelity of these 12 measurements increased from  0.98 $\pm$ 0.01 for uncorrected measurements to 0.994 $\pm$ 0.009 for corrected measurements, bringing the value closer to the optical case of F = 0.996 $\pm$ 0.007 for this measurement subset.\\

\begin{figure*}
  \includegraphics[width=\linewidth]{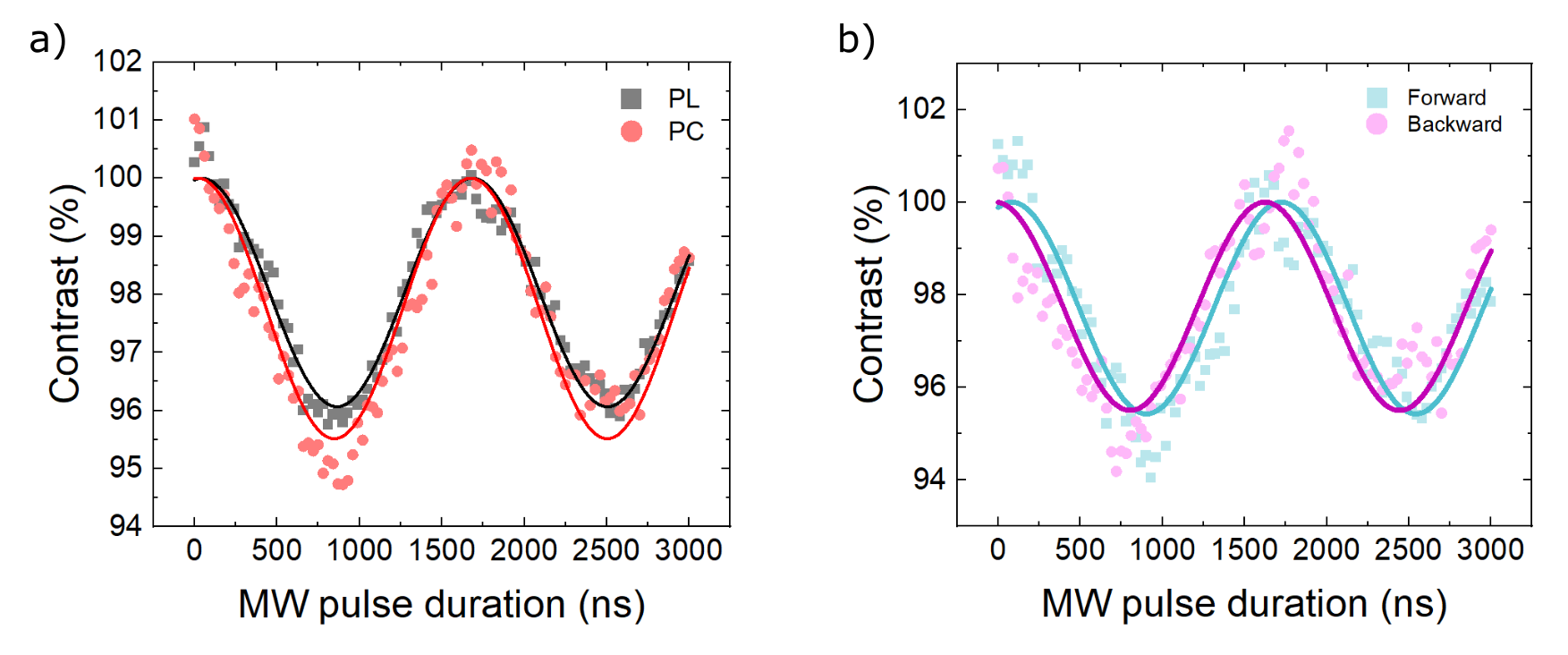}
  \caption{\textbf{a)} Y-Rabi of PC-RPQST (red) and PL-RPQST (black) for the state ($\theta$, $\phi$) = ($15.37^{\circ}$, $235^{\circ}$) fitted by sinusoids. \textbf{b)} The forward (blue) and backward (pink) $PC$ y-Rabi measurements. The phase shift between $PL$- and forward $PC$-Rabi is $\sim 20^{\circ}$. }
  \label{meas_BS}
\end{figure*}

As an alternative to the previously presented correction method, we tested a backsweep method. In this method, each Rabi measurement is performed in pairs of sweeps: one with increasing $\tau$ and the next in reverse (with decreasing $\tau$ - the backsweep). Averaging them cancels the phase shift for every consecutive pair of measurements, meaning that even systems with a response that varies at timescales of a few experiment runs can be accurately represented. Using the backsweep method, the state fidelity of the aforementioned 13 measurements increases to 0.997 $\pm$ 0.002.  The method is illustrated in \textbf{figure \ref{meas_BS}} for a single PC-RPQST measurement (the Y-Rabi shown in \textbf{figure \ref{meas_QST}}). In particular, \textbf{figure \ref{meas_BS}a)} shows the $PL$-Rabi together with the $PC$-Rabi that was detected using the backsweep method. Their phase difference is $\sim$ 2$^{\circ}$, which is within the noise level that is determined by the number of repetitions. \textbf{Figure \ref{meas_BS}b)} shows the forward and backward sweeps that make up the PC-signal. The forward sweep shows a phase lag of $\sim 20^{\circ}$ compared to the $PL$-Rabi while the backward sweep shows a phase lead of approximately equal size. Added together, these phases are canceled out. Due to the increased state fidelity, this method of correcting for the $PC$ phase lag was chosen and used in all of the following discussions.\\

The effect of current readout delay has another consequence besides introducing the discussed systematic error. Namely, it is putting an upper limit to the speed of tomography. From the analysis of the PC time traces (see \textbf{Supplementary Information}), we conclude that the current readout delay originates from the mechanism of charge collection from the diamond. Therefore, we propose  that the key to solving the issue of the delay could lie in further material and contact optimization. This would need to be confirmed with a dedicated study.

\subsection{Results for PC-RPQST}
\begin{figure*}
  \includegraphics[width=\linewidth]{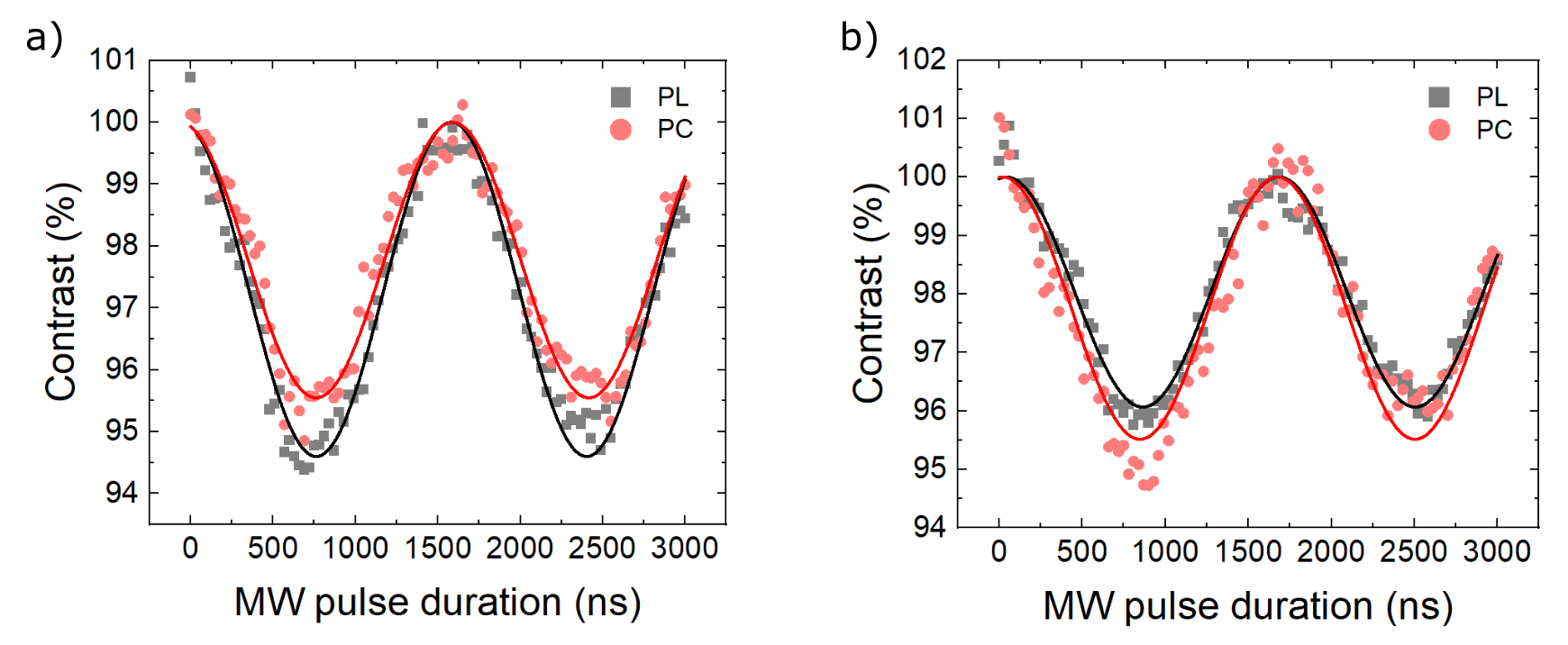}
  \caption{PC- (red) and PL-RPQST (black) for the state ($\theta$, $\phi$) = ($15^{\circ}$, $235^{\circ}$). The experimental data is fitted and phase angles of \textbf{a)} x- and \textbf{b)} y-Rabi are extracted. From these parameters, in this particular measurement, the electron spin state was reconstructed with a state fidelity of 0.99991 (PL-RPQST) and 0.9998 (PC-RPQST). }
  \label{meas_QST}
\end{figure*}

In the PC-RPQST measurements, the backsweep method was used to mitigate for the $PC$ delay. \textbf{Figure \ref{meas_QST}a)} and \textbf{b)} show an example of raw data and sinusoidal fits of one of these measurements, where the Rabi oscillations around the x-axis are on the left and the oscillations around the y-axis on the right. The tomographed state was prepared with ($\theta$, $\phi$) = ($15.37^{\circ}$, $235^{\circ}$) and using PC-RPQST, it was reconstructed as ($15.13^{\circ}$, $241.47^{\circ}$) resulting in a state fidelity of 0.9998 for this particular measurement. The state fidelity of the corresponding PL-RPQST measurement was 0.99991.\\

\begin{figure}
  \includegraphics[width=\linewidth]{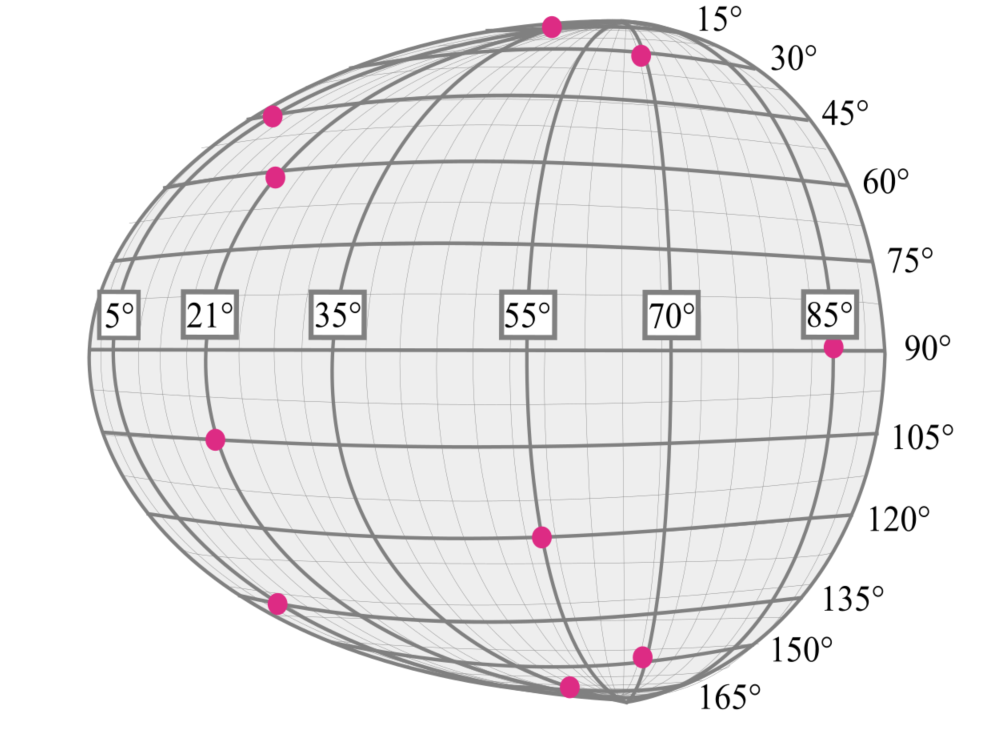}
  \caption{Electron spin states that were used in Rabi measurements, projected to the first quadrant. Each pink dot represents a state that was prepared and tomographed at least twice. The azimuthal angle $\theta$ is indicated next to the Bloch sphere quarter, while the polar angle $\phi$ is plotted on top of the sphere.}
  \label{states}
\end{figure}

We performed 21 measurements of 10 different electron spin states that are depicted in \textbf{figure \ref{states}} with $\phi$ recalculated to the first quadrant as allowed by the spherical symmetry (for exact state angles, see \textbf{Supplementary Information}). These states are representative for the entire Bloch sphere and we found an average state fidelity of $0.995 \pm 0.006$ for PC-RPQST and $0.994 \pm 0.009$ for the simultaneous optical measurements. This is similar to the earlier reported value of $0.995 \pm 0.005$ for RPQST using the conventional $PL$ readout. \cite{ourQST} Hence, we have demonstrated that adopting the electrical readout does not reduce the state fidelity of the tomography methodology.\\

Upon closer inspection of the reconstructed states (see Table 1 in \textbf{Supplementary Information}), errors ($\Delta \theta$, $\Delta \phi$) in the measured state angles $(\theta, \phi)$ can be divided into random and systematic errors (e.g. sample drift, inhomogeneities in the MW control field, etc.), which are largest for states near the Bloch sphere equator and smaller at the poles as is expected from \cite{ourQST}. By canceling out the effect of the systematic errors, we can calculate the state fidelity that is possible to attain using PC-RPQST with optimized measurement setup. To do so, we look at the average and standard deviation of $\Delta \theta$ and $\Delta \phi$. Taking into account only standard deviation, systematic errors are not considered and we find an optimized state fidelity of $0.998 \pm 0.001$, which may be attained by hardware and software improvements such as suppression of decoherence by, e.g., using dynamical decoupling or numerically designed pulses.

\section{Conclusion} \label{conclusion}
In conclusion, we have carried out the quantum state tomography using PC-RPQST methodology implemented by photoelectric detection of the NV electron spin state, where the pulse sequence was adapted for the electrical readout by repeating segments of the sequence inside low frequency envelopes. A delay in photocurrent resulted in a phase shift of the $PC$-Rabi signals compared to their optical counterparts, which was detrimental for the PC-RPQST state fidelity. Several methods to correct for this effect are discussed. From these, the backsweep method is shown to be most successful. This means that every Rabi measurement was performed in pairs of sweeps, one normal (increasing Rabi period - forward sweep) and one in the opposite direction (decreasing Rabi period - backsweep). Using this method, the state fidelity of PC-RPQST is almost identical to the simultaneous optical measurements. In particular, for 21 measurements of 10 different electron spin states, the state fidelity was 0.995 $\pm$ 0.006 for PC-RPQST and 0.994 $\pm$ 0.009 optically. However, the measurement speed is the remaining issue and needs to be addressed for the PC-RPQST methodology to become competitive.\\

This work demonstrates that the photocurrent readout is a useful tool to measure spin states with high-fidelity. Further improvements in response speed are expected to contribute to more precise measurements without the need for phase delay compensation. Such improvements could be achieved upon precise determination of the PC delay mechanism and its mitigation by optimizing the diamond material properties and electrical contacts, which is the subject of follow-up research.   \\

\textbf{Acknowledgments} 
The research was co-financed by the Research Foundation-Flanders via FWO file number 11O1625N. We also acknowledge the support from Eu projects 101113901 and 101113983 (QU-Test and QU-Pilot projects), Grant Agreement No. 101135699 (SPINUS), Grant Agreement No. 101135359 (C-QUENS) and GACR project 24-12984S, FWO Project No. G0D1721N and EOS CHEQs Project No. 40007526. Lastly, we thank A. Shukla for providing the original idea of Rabi Quantum State Tomography, of which this work demonstrates an extension. The authors declare no conflicts of interest.

\bibliography{ref2}
\bibliographystyle{apsrev4-2}
\end{document}